\documentclass[preprint,12pt]{elsarticle}

\def\lbldef#1#2{\expandafter\gdef\csname #1\endcsname {#2}}

\def\href#1#2{#2}

\newcommand{\be}{\begin{equation}}
\newcommand{\ee}{\end{equation}}
\newcommand{\bea}{\begin{eqnarray}}
\newcommand{\eea}{\end{eqnarray}}
\newcommand{\hunit}{\mathrm{km \ s^{-1} \ Mpc^{-1}}}
\newcommand{\hii}{HII}
\newcommand{\hb}{\mathrm{H}\beta}

\usepackage{color}
\usepackage{graphicx}
\usepackage{amssymb}
\journal{Physics of the Dark Universe}

\begin{document}

\begin{frontmatter}

\title{Testing the $R_{\rm h}=ct$ Universe Jointly with the \\
Redshift-dependent Expansion rate and \\
Angular-diameter and Luminosity Distances}

\author[1]{Hao-Yi Wan\footnote{E-mail: haoyimail@gmail.com}}
\author[2]{Shu-Lei Cao\footnote{E-mail: shulei@ksu.edu}}
\author[3]{Fulvio Melia\footnote{John Woodruff Simpson Fellow. E-mail: fmelia@email.arizona.edu}} 
\author[4]{Tong-Jie Zhang\footnote{E-mail: tjzhang@bnu.edu.cn}}

\address[1]{Department of Astronomy, Beijing Normal University, Beijing 100875, China}
\address[2]{Department of Astronomy, Beijing Normal University, Beijing 100875, China}
\address[3]{Department of Physics, The Applied Math Program, and \\ 
Department of Astronomy, The University of Arizona, AZ 85721, USA}
\address[4]{Department of Astronomy, Beijing Normal University, Beijing 100875, China; \\
and Institute for Astronomy Science, Dezhou University, Dezhou 253023, China}

\begin{abstract}
We use three different data sets, specifically $H(z)$ measurements from cosmic
chronometers, the HII-galaxy Hubble diagram, and reconstructed quasar-core
angular-size measurements, to perform a joint analysis of three flat
cosmological models: the $R_{\rm h}=ct$ Universe, $\Lambda$CDM, and
$w$CDM. For $R_{\rm h}=ct$, the 1$\sigma$ best-fit value of the Hubble
constant $H_0$ is $62.336\pm1.464$ $\hunit$, which matches previous
measurements ($\sim 63$ $\hunit$) based on best fits to individual data
sets. For $\Lambda$CDM, our inferred value of the Hubble constant,
$H_0=67.013\pm2.578$ $\hunit$, is more consistent with the {\it Planck}
optimization than the locally measured value using \mbox{Cepheid} variables,
and the matter density $\Omega_{\rm m}=0.347\pm0.049$ similarly coincides
with its {\it Planck} value to within 1$\sigma$. For $w$CDM, the optimized
parameters are $H_0=64.718\pm3.088$ $\hunit$, $\Omega_{\rm m}=0.247\pm0.108$
and $w=-0.693\pm0.276$, also consistent with {\it Planck}. A direct
comparison of these three models using the Bayesian Information Criterion
shows that the $R_{\rm h}=ct$ universe is favored by the joint analysis
with a likelihood of $\sim 97\%$ versus $\lesssim 3\%$ for the other two
cosmologies.
\end{abstract}

\end{frontmatter}

\section{Introduction}
\label{sec:intro}
We are witnessing a surge in activity finding new ways to test the
observational signatures of competing cosmological models. These
include tests based on the measurement of the expansion rate $H(z)$
with cosmic chronometers \citep{Jimenez2002,Moresco2012}
and baryon acoustic oscillations (BAO) \citep{Blake2012},
the HII Hubble diagram constructed from the luminosity distance to HII
galaxies \citep{Melnick1987,Melnick1988,Fuentes-Masip2000,Melnick2000,Bosch2002,Telles2003,Siegel2005,Bordalo2011,Plionis2011,Mania2012,Chavez2012,Terlevich2015}, and the
angular-diameter distance inferred from the angular-size of compact cores
in quasars \citep{Gurvits1999,Cao2015,Cao2017a,Cao2017b}. These,
and other recent observations, are revealing some tension between the
predictions of the standard model $\Lambda$CDM and the actual measurements,
including the value of the Hubble constant, $H_0$, measured locally with
Cepheid variables and at high redshift with {\it Planck} \citep{Wei2017};
$\sigma_8$ \citep{Percival2009,Macaulay2013,Pavlov2014,Alam2016} and the
structure growth rate measured by 2dFGRS \citep{Peacock2001}, VVDS
\citep{Guzzo2008}, quasar clustering \citep{Ross2007,daAngela2008,Viel2004},
and peculiar velocity studies \citep{Davis2011,Hudson2012}]; the anomalously
early appearance of galaxies \citep{Melia2014b} and massive quasars
\citep{Melia2013a,Melia2015b}; and the redshift-dependent halo-mass
distribution function \citep{Steinhardt2016,Yennapureddy2018}.

There is therefore some motivation to consider new \mbox{physics}, or alternative
cosmological models. Thus far, arguably the most successful competitor to
$\Lambda$CDM has been the $R_{\rm h}=ct$ universe \citep{Melia2007,Melia2009,Melia2012a}.
In this model, (i) the observed constraint of our gravitational horizon, i.e.,
$R_{\rm h}(t_0)=ct_0$, is upheld for all cosmic time $t$ which, however,
would not be entirely consistent with $\Lambda$CDM, or any other
cosmological model we know of \citep{Melia2007}, (ii) the expansion rate,
$\dot{a}$, is constant, where $a$ is the scale factor and the overdot
signifies differentiation with respect to $t$, and (iii) the pressure
$p$ and energy density $\rho$ satisfy the zero active mass condition in
general relativity, $p=-\rho/3$, which appears to be a requirement for
the simplified form of the metric in the Friedmann-Robertson-Walker (FRW)
spacetime \citep{Melia2016b}.

But up until now, all of the tests carried out for this model have been
based on the use of individual data sets (see Table~1 in ref.~\cite{Melia2017a}).
In this paper, for the first time,
we consider a joint analysis and test of this model using three very different
kinds of measurement, including the redshift-dependent expansion rate, and the
angular-diameter and luminosity distances. Other recent model comparisons have
shown that the data favor predominantly the standard and $R_{\rm h}=ct$
cosmologies \citep{Wei2017,Melia2018}, so in this paper we streamline our
comparative analysis by ignoring other models sometimes used for model
selection, including Einstein-de Sitter and the Milne universe.

We shall introduce the measurements of $H(z)$ based on cosmic chronometers,
the HII-galaxy Hubble diagram, and the angular size of quasar cores in
Section~\ref{data}. Then, in Section~\ref{model}, we shall compare the
predicted expansion rates and distances in these three models with the
data. The joint analysis and results of our comparison will be presented
in Section~\ref{analysis} and we shall discuss the outcome in
Section~\ref{discussion}. Section~\ref{conclusion} presents our conclusions
and future prospects.

\section{Data}
\label{data}
\subsection{$H(z)$ measurements}
There are two predominant methods used to measure $H(z)$: cosmic chro\-nometers
\citep{Jimenez2002,Moresco2012} and the baryon acoustic oscillation
(BAO) scale \citep{Blake2012}. The former is based on a direct
determination of the relative ages of adjacent galaxies, and is model independent.
As discussed more extensively in \cite{Melia2013c}, however, the $H(z)$ measurements
extracted from the BAO scale often require the pre-assumption of a specific
cosmological model in order to separate the ever-present redshift space
distortions from the actual redshift width and position of the BAO peak.
In addition, this method \mbox{depends} on how ``standard rulers'' evolve with redshift,
rather than how cosmic time changes locally with $z$. Unfortunately, the expansion
rates measured in these different ways are sometimes combined to produce an overall
$H(z)$ versus $z$ diagram, but they cannot be used to test different models because,
as noted, the BAO approach necessarily adopts a particular cosmology and is
therefore model-dependent. Thus, unlike other recent attempts at using measurements
of $H(z)$ for cosmological work \citep{Cao2018}, we use only cosmic chronometers
in this paper, as listed in Table~1 of ref.~\citep{Ruan2019} (assembled from various
published sources cited in that work).

\subsection{HII-Galaxy Hubble Diagram}
We use a total sample of 156 sources (25 high-$z$ \hii~galaxies, 107 local \hii~galaxies, 
and 24 giant extragalactic \hii~regions), assembled by \cite{Terlevich2015}. 
A detailed description of these objects, their relevant properties and the methodology used
to derive them, was given in refs.~\citep{Terlevich2015,Erb2006,Hoyos2005,Maseda2014,Masters2014,Chavez2014}.
With these data,\footnote{We also point out that an incorrect reference was previously given for the 
H$\beta$ flux of the three sources Q2343-BX660, HoyosD2-5, and HoyosD2-1 in Table~1 of ref.~\cite{Wei2016}. 
The correct attribution should have been ref.~\cite{Terlevich2015}.}
the H$\beta$ luminosity may be calculated according to the expression
\begin{equation}
L(\hb) = 4 \pi  d_L^2(z) F(\hb)\;,
\end{equation}
where $d_L$ and $F(\mathrm{H}\beta)$ are the cosmological model-dependent luminosity
distance at redshift $z$ and the reddening-corrected $\hb$ flux, respectively.

The emission-line luminosity versus ionized-gas velocity dispersion correlation
($L - \sigma$) reads \citep{Chavez2012,Chavez2014,Terlevich2015}
\begin{equation}
\log L(\hb)=\alpha \log \sigma(\hb)+\kappa\;,
\end{equation}
where $\alpha$ is the slope of the function and $\kappa$ is a constant representing
the logarithmic luminosity at $\log \sigma(\hb)=0$. This empirical correlation for
$L(\hb)$ has been used as a luminosity indicator for cosmology
(e.g., \cite{Chavez2012,Terlevich2015}), but the cosmological parameters
$\alpha$ and $\kappa$ are unknown and must be considered as free parameters
to be optimized simultaneously with those of the cosmological model.

The expression for the observed distance modulus of an HII galaxy was deduced by 
ref.~\citep{Chavez2016}, based on their fittings to their observed $L - \sigma$ relation. 
Combining their Equations~(1) and (2), one may write the distance modulus as
\begin{equation}\label{eq:HII}
\mu_{\rm obs}=2.5\left[\kappa +\alpha \log \sigma(\hb) - \log F(\hb)\right]-100.2 \ ,
\end{equation}
and its error $\sigma_{\mu_{\rm obs}}$ can be calculated by error propagation,
i.e.,
\begin{equation}\label{sigma}
\sigma_{\mu_{\rm obs}}=2.5\sqrt{(\alpha \sigma_{\log \sigma})^{2}+(\sigma_{\log F})^{2}} \;,
\end{equation}
where $\sigma_{\log \sigma}$ and $\sigma_{\log F}$ are the $1\sigma$ uncertainties
of $\log \sigma(\hb)$ and $\log F(\hb)$, respectively.

By comparison, the theoretical distance modulus $\mu_{\rm th}$ is defined as
\begin{equation}
\mu_{\rm th}\equiv5 \log\left[\frac{d_{L}(z)}{\rm Mpc}\right]+25\;.
\end{equation}
This is the expression we compare with Equation~(3) to determine the quality of
the fit in each model.

\begin{table}
\centering
\small
\caption{Quasar-Core Data}\label{tab:Qus}
\setlength{\tabcolsep}{10mm}{
\begin{tabular}{lcc}
\hline\hline
\\
 $z$ & $\theta_{\rm core}$ & $\sigma_{\theta_{\rm core}}$ \\
     & (mas)               & (mas) \\
\\
\hline
\\
0.5180 & 2.3990 & 0.6643 \\
0.6320 & 1.9330 & 0.3737 \\
0.7255 & 1.4540 & 0.2243 \\
0.8370 & 1.7300 & 0.4597 \\
0.9365 & 1.4740 & 0.4087 \\
1.0325 & 1.5440 & 0.2782 \\
1.1045 & 1.3990 & 0.4172 \\
1.1715 & 1.2770 & 0.4492 \\
1.2450 & 1.5440 & 0.3838 \\
1.2920 & 1.2000 & 0.4595 \\
1.3290 & 1.3870 & 0.5055 \\
1.3900 & 0.7580 & 0.3794 \\
1.4525 & 1.3300 & 0.2510 \\
1.5245 & 1.2800 & 0.2917 \\
1.6255 & 1.4310 & 0.3751 \\
1.7325 & 1.4640 & 0.3213 \\
1.8120 & 1.1360 & 0.4476 \\
1.9490 & 1.3960 & 0.2804 \\
2.0785 & 1.5960 & 0.5076 \\
2.4960 & 1.4560 & 0.1858 \\
\\
\hline\hline
\end{tabular}}
\end{table}

\subsection{Quasar cores}
Jackson et al. \cite{Jackson2006} assembled a sample of ultra-compact radio
sources, extracted from an old 2.29 GHz VLBI survey of \cite{Preston1985} and
additions by \cite{Gurvits1994}. A mixed population of AGNs and quasars, however,
makes it difficult to disentangle systematic differences from true cosmological
variations. Fortunately, this issue has been overcome by a recent introduction of
an additional luminosity restriction applied to these sources, used in conjunction
with the constraint on the spectral index $\epsilon$ that was used earlier by
\cite{Gurvits1999}. The exclusion of sources with low luminosities $L$ could
mitigate the dependence of the intrinsic core size on $L$ and redshift $z$
\citep{Vishwakarma2001}.  The physical core size $\ell_{\rm core}$
strongly depends on luminosity both at the low and high ends \citep{Cao2017a,Cao2017b}.
It appears that a sub-sample of \cite{Jackson2006} with spectral index $-0.38<\epsilon<0.18$
and intermediate luminosity $10^{27}$ W/Hz $<L<10^{28}$ W/Hz has a reliable standard linear
size. Here, to partially minimize an additional degree of scatter that would otherwise
appear using the individual data points, we bin these sources into groups of 7 and select
the median value in each bin to represent the angular size $\theta_{\rm core}$
\citep{Santos2008}. The resulting 20 data points are listed in Table \ref{tab:Qus},
along with their 1$\sigma$ errors estimated assuming Gaussian variation within each
bin. The theoretical angular size of the compact quasar core is given as
\begin{equation}\label{eq:quasar}
\theta^{\rm th}_{\rm core}=\frac{\ell_{\rm core}}{d_A(z)},
\end{equation}
where $d_A$ is the model-dependent angular diameter distance.

\section{Models and Observational Signatures}
\label{model}
In this paper, we consider three flat cosmological models:
\begin{enumerate}
  \item The $R_{\rm h}=ct$ Universe (an FRW cosmology with zero active mass and
  equation of state, EoS, $\rho+3p=0$ \cite{Melia2007,Melia2016b,Melia2017b}).
  In this model, the Hubble parameter $H(z)$, luminosity distance $d_L$ and
  angular-diameter distance $d_A$ are given, respectively, as
      \be\label{Hz1}
      H(z)=H_0(1+z),
      \ee
      \be\label{dl1}
      d_L(z)=\frac{c}{H_0}(1+z)\ln(1+z),
      \ee
      and
      \be\label{da1}
      d_A(z)=\frac{c}{H_0}\frac{1}{1+z}\ln(1+z),
      \ee
      where $c$, $z$ and $H_0$ are the speed of light, redshift and Hubble constant, respectively.
  \item Flat $\Lambda$CDM, with matter density parameter $\Omega_{\rm m}$, a constant dark energy
      (DE) density parameter $\Omega_{\Lambda}$ and EoS $w_{\Lambda}=-1$. For this model,
      \be\label{Hz2}
      H(z)=H_0\sqrt{\Omega_{\rm m}(1+z)^3+\Omega_{\Lambda}}
      \ee
      (ignoring the insignificant contribution from radiation in the local Universe),
      \be\label{dl2}
      d_L(z)=\frac{c}{H_0}(1+z)\int^z_0\frac{dz'}{\sqrt{\Omega_{\rm m}(1+z')^3+\Omega_{\Lambda}}},
      \ee
      and
      \be\label{da2}
      d_A(z)=\frac{c}{H_0}\frac{1}{1+z}\int^z_0\frac{dz'}{\sqrt{\Omega_{\rm m}(1+z')^3+\Omega_{\Lambda}}}.
      \ee
  \item Flat $w$CDM, which differs from $\Lambda$CDM by having a variable DE EoS, so that
      \be\label{Hz3}
      H(z)=H_0\sqrt{\Omega_{\rm m}(1+z)^3+\Omega_{\rm de}(1+z)^{3(1+w_{\rm de})}},
      \ee
      \be\label{dl3}
      d_L(z)=\frac{c}{H_0}(1+z)\int^z_0\frac{dz'}{\sqrt{\Omega_{\rm m}(1+z')^3+\Omega_{\rm de}(1+z')^{3(1+w_{\rm de})}}},
      \ee
      and
      \be\label{da3}
      d_A(z)=\frac{c}{H_0}\frac{1}{1+z}\int^z_0\frac{dz'}{\sqrt{\Omega_{\rm m}(1+z')^3+\Omega_{\rm de}(1+z')^{3(1+w_{\rm de})}}}.
      \ee
\end{enumerate}

\section{Joint Analysis and Results}
\label{analysis}
We begin by first describing our model selection procedure, based on the Bayes
Information Criterion (BIC). This approach maximizes the joint likelihood function for all
of the fitting parameters, which can be viewed as a function to be maximized, or as a
multiplicative factor modifying an assumed Bayesian prior. Since the uncertainties for both
the $H(z)$ measurements and the angular-sizes of quasar cores $\theta_{\rm core}$ are known,
based on flat Bayesian priors, their likelihood functions can be presented as follows:
\be\label{likHz}
\mathcal{L}_{H(z)}\propto\exp\left[-\sum_i\frac{\left[H_{\rm obs,i}-H_{\rm th}(z_i)\right]^2}
{2\sigma^2_{H_{\rm obs,i}}}\right],
\ee
and
\be\label{likQus}
\mathcal{L}_{\theta_{\rm core}}\propto\exp\left[-\sum_i\frac{\left[\theta_{\rm core,i}-
\theta^{\rm th}_{\rm core}(z_i)\right]^2}{2\sigma^2_{\theta_{\rm core,i}}}\right],
\ee
where $H_{\rm obs,i}$ and $\theta_{\rm core,i}$ are the observed expansion rates and
quasar core angular-sizes, respectively, along with their 1$\sigma$ uncertainties
(see Table~\ref{tab:Qus}). Also, $H_{\rm th}(z_i)$ and $\theta^{\rm th}_{\rm core}(z_i)$
are the theoretical values of $H(z)$ and $\theta_{\rm core}$ evaluated at redshift $z_i$.
For the HII galaxies, however, the error $\sigma_{\mu_{\rm obs}}$ depends on the value
of $\alpha$ (see Eq.~4), so the likelihood function is instead
\be\label{likHII}
\mathcal{L}_{\rm \hii} = \prod_{i}
\frac{1}{\sqrt{2\pi}\,\sigma_{\mu_{\rm obs}, i}}\;\times
\exp\left[-\,\frac{\left(\mu_{{\rm obs},i}-\mu_{\rm
      th}(z_i)\right)^{2}}{2\sigma^{2}_{\mu_{\rm obs}, i}}\right] .
\ee
The joint likelihood function used in this analysis is therefore\break 
\noindent $\mathcal{L}_{\rm joint} =\mathcal{L}_{H(z)}\mathcal{L}_{\theta_{\rm core}}\mathcal{L}_{\rm \hii}$.

\begin{figure}
\begin{center}
\includegraphics[width=1.0\linewidth]{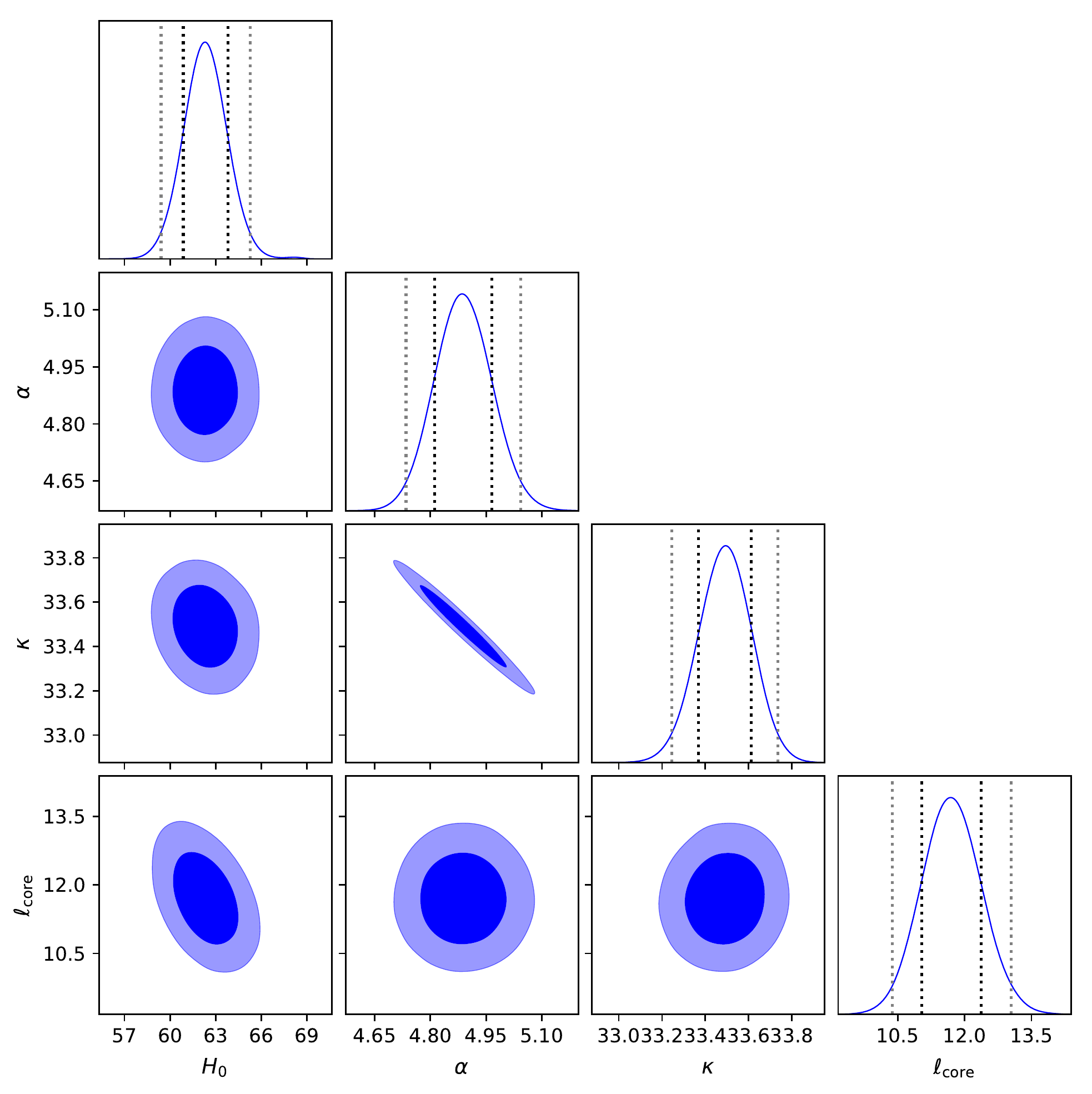}
\end{center}
\caption{1-2$\sigma$ confidence regions of the parameters $H_0$, $\alpha$,
               $\kappa$, and $\ell_{\rm core}$ for the $R_{\rm h}=ct$ Universe,
               along with their probability distributions.}\label{fig1}
\end{figure}

\begin{figure}
\begin{center}
\includegraphics[width=1.0\linewidth]{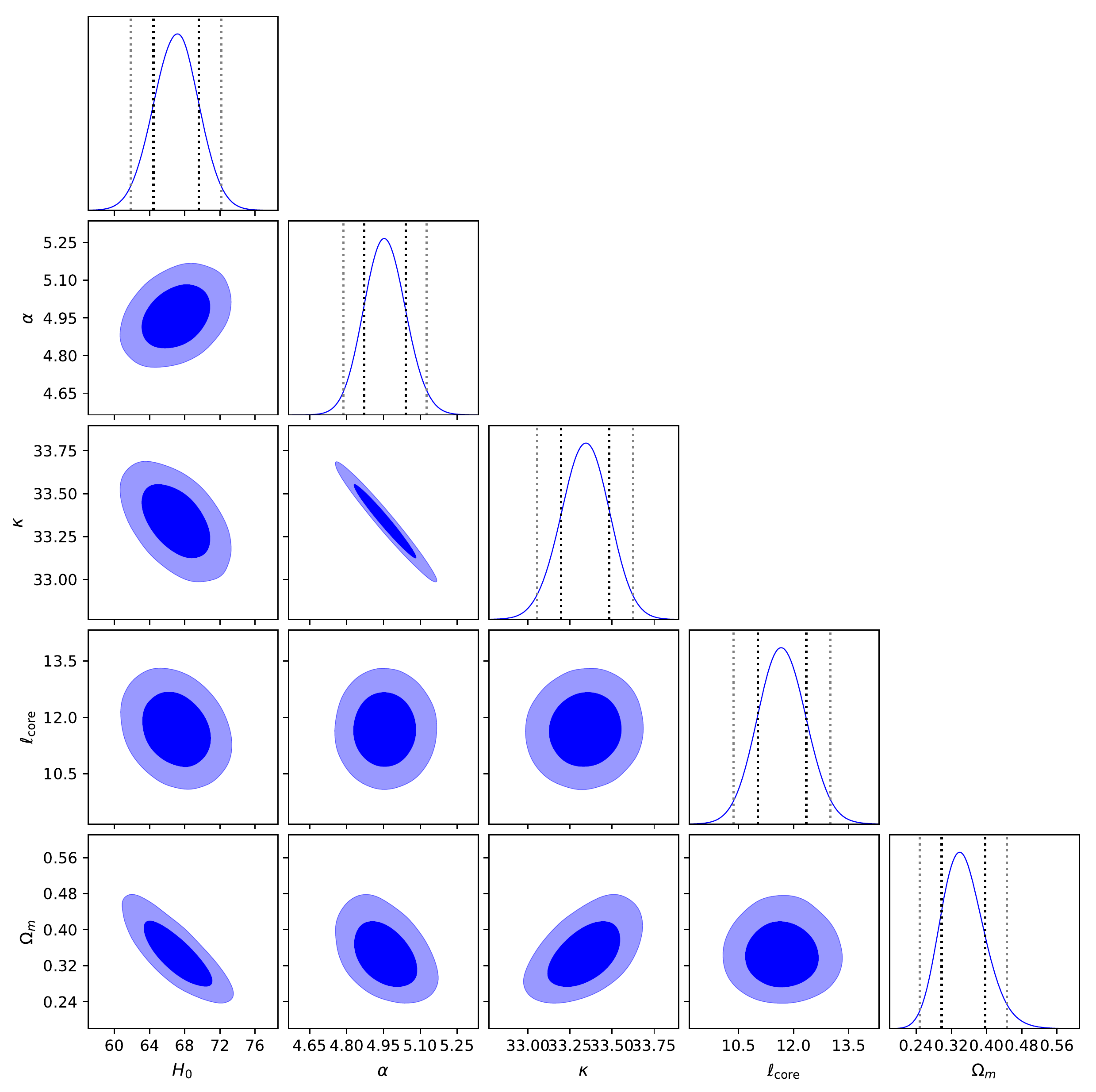}
\end{center}
\caption{1-D probability distributions and 2-D regions with 1-2$\sigma$
               contours for the parameters $H_0$, $\alpha$, $\kappa$,
               $\ell_{\rm core}$, and $\Omega_{\rm m}$ for flat $\Lambda$CDM.}\label{fig2}
\end{figure}

From the resulting model-specific likelihood function
$\mathcal{L}_{\rm joint}$, which may also be viewed as a Bayesian posterior, we
optimize the parameters by maximizing over their joint space. Such a procedure can
be biased, so to avoid this, we also consider the shape of the function
$\mathcal{L}_{\rm joint}$ near its maximum, as described below.

\begin{figure}
\begin{center}
\includegraphics[width=1.0\linewidth]{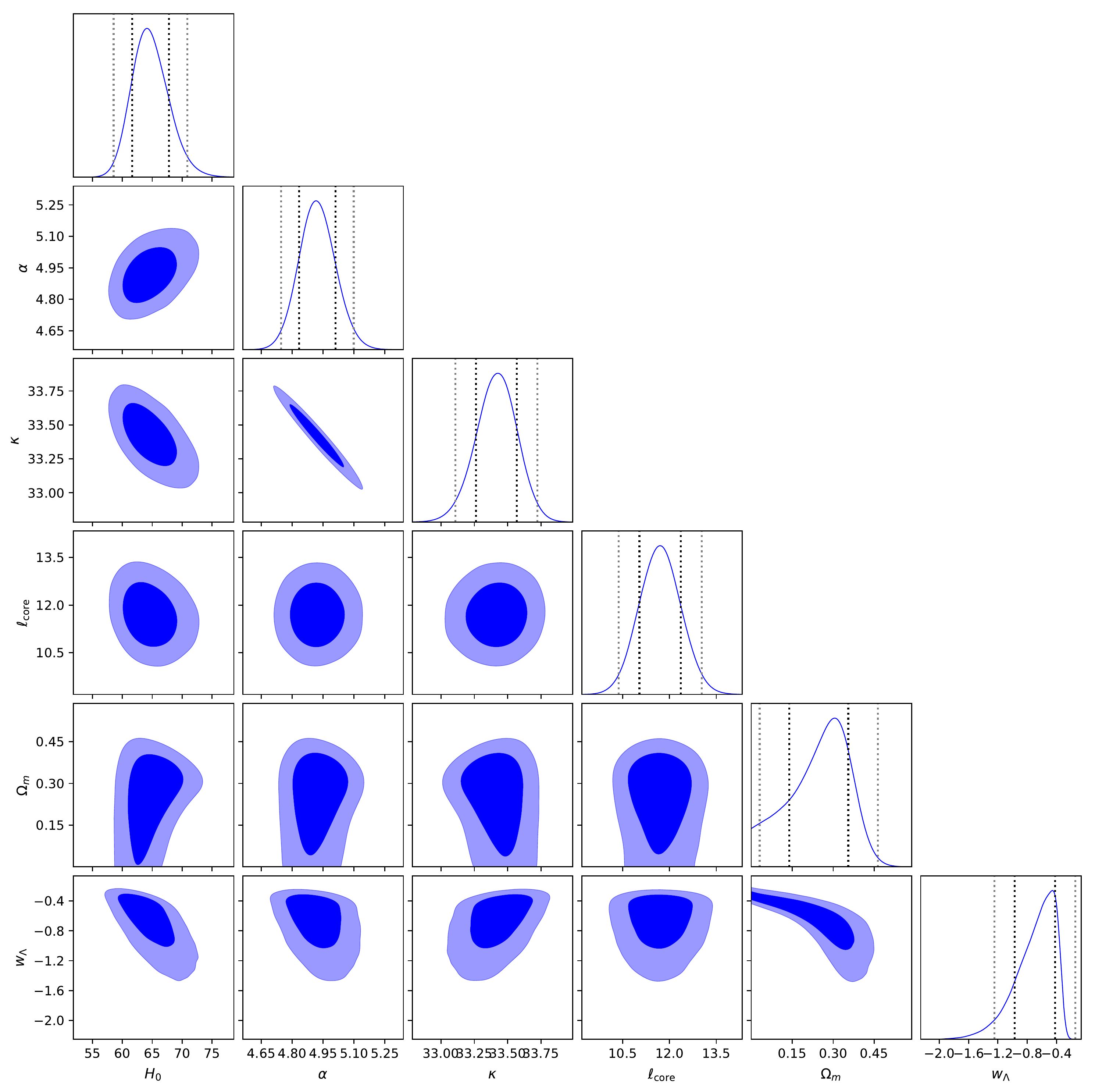}
\end{center}
\caption{1-2$\sigma$ constraints on $H_0$, $\alpha$, $\kappa$, $\ell_{\rm core}$,
               $\Omega_{\rm m}$, and $w_{\Lambda}$ for the flat $w$CDM model.}\label{fig3}
\end{figure}

\begin{figure}
\begin{center}
\includegraphics[width=1.0\linewidth]{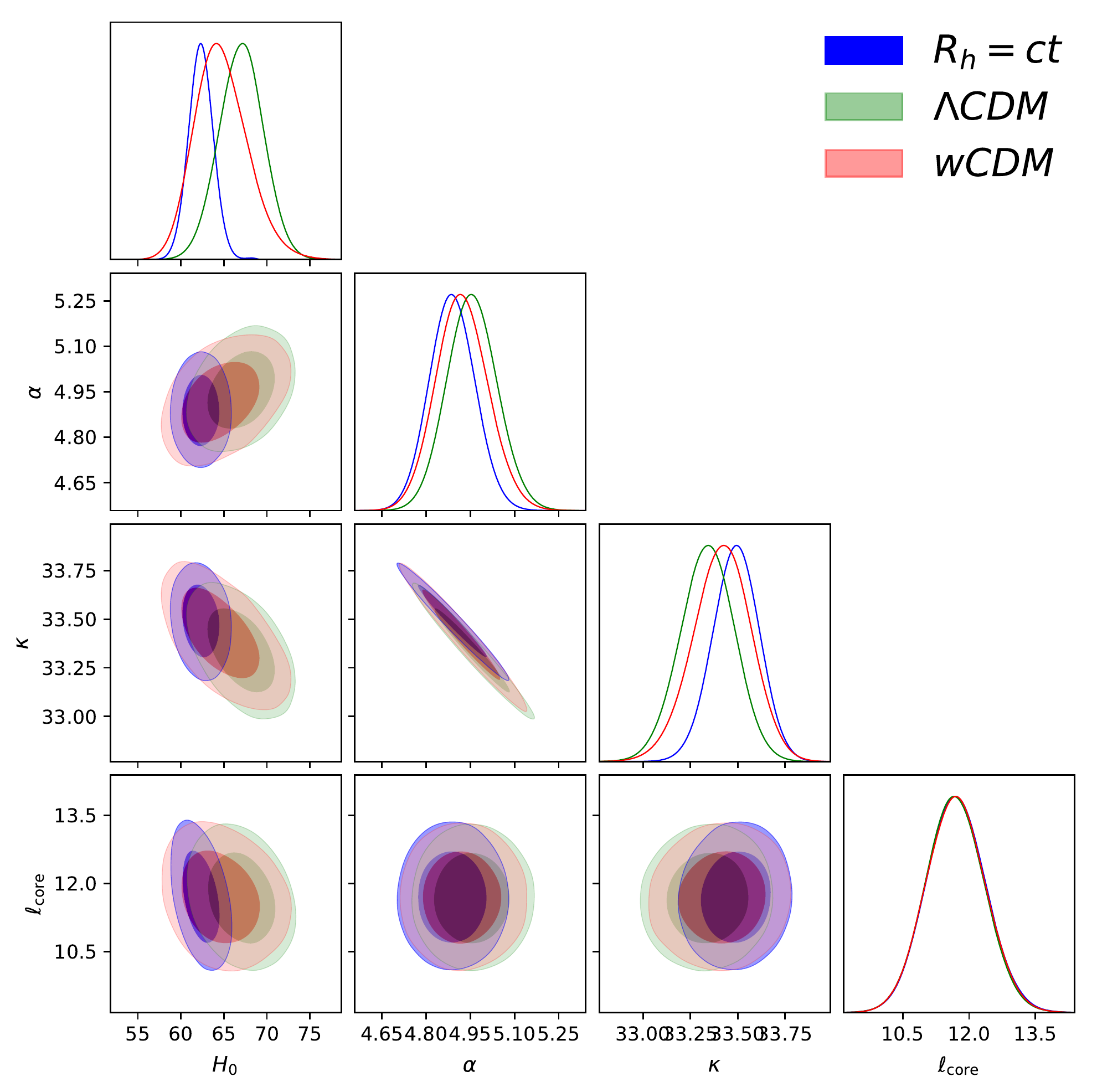}
\end{center}
\caption{A comparison of the probability distributions and 1-2$\sigma$ confidence regions
               for $H_0$, $\alpha$, $\kappa$, and $\ell_{\rm core}$, for all three models.}\label{fig4}
\end{figure}

In addition to optimizing these parameters by maximizing $\mathcal{L}_{\rm joint}$,
we treat this joint likelihood function in a Bayesian fashion as an unnormalized
probability density function on the joint parameter space, including the coefficients
$\alpha$ and $\kappa$, and employ the pure-Python implementation of an affine invariant
Markov chain Monte Carlo (MCMC) ensemble sampler called emcee \citep{Foreman-Mackey2013}
to generate a large random sample of points from this space, distributed according to
the probability density function. This procedure allows us to determine the 1-D probability
distributions and 2-D confidence regions with 1-2$\sigma$ contours for all the free
parameters in each model, and these are shown in Figs. \ref{fig1}-\ref{fig3}, along with
a comparison for all three models in Fig. \ref{fig4}.

In statistics, the Bayesian information criterion (BIC) is commonly used for model
selection among a finite set of models. The BIC is formally defined as
\be\label{bic}
\mathrm {BIC} =k\ln(n)-2\ln(\mathcal{L}_{\rm joint}),
\ee
where $k$ is the number of free parameters and $n$ is the number of data points.
According to \cite{Melia2013c}, in a comparison between three cosmologies, model
$\alpha$ has a likelihood
\be
\label{eq:prob}
\mathcal{L}_{\alpha}=\frac{\exp(-\rm BIC_{\alpha}/2)}{\exp(-\rm BIC_{1}/2)+\exp(-\rm BIC_{2}/2)+\exp(-\rm BIC_{3}/2)}
\ee
of being the best choice.
The best-fit values of the free parameters in each model, the BIC value for the optimized
fit, and the corresponding probabilities, are listed in Table \ref{tab:jonit}.

The use of the BIC is especially motivated when the samples are very large
\cite{Melia2013c}, which approximates the computation of the (logarithm of the) `Bayes
factor' for deciding between models \cite{Schwarz1978,Kass1995}.  It is important to
emphasize here that in the limit of large sample size~$n$, particularly when $n\gg k$,
the posterior distribution typically becomes increasingly peaked and Gaussian in shape.
Kass \& Raftery \cite{Kass1995} used this property to justify the BIC, which is often
cited in the statistics literature.  As long as the parameters have a unimodal
distribution that is roughly Gaussian, which turns~out to be the case for the analysis
in this paper, the Bayes factor between competing models can be calculated to high
accuracy from the ratio of their respective (maximized) likelihoods. In the Kass
\& Raftery argument, the definite integrals of increasingly peaked integrands are
approximated using Laplace's method, similar to Stirling's approach of calculating
an asymptotic approximation to $n!$ when $n\gg 1$, as seen in his well-known formula.
For the samples we use here, $ n\gg 1$, which is deep in the asymptotic regime.

From Equation~(20), we thus recognize an obvious Bayesian interpretation:
$\exp(-{\rm BIC}_\alpha/2)$ is a large-sample $(n\to\infty)$ approximation to an
integral over the joint parameter space of model~$\alpha$, of its joint
likelihood function $\mathcal{L}_{\rm joint}$. This integral could be evaluated
by brute-force numerical calculations in low-dimensional cases. But to see that
our approximation here is reasonable, suppose that $\mathcal{L}_{\rm joint}$ is
Gaussian near its maximum, where it equals $\mathcal{L}^*_{\rm joint}$, and the
$k$~parameters, called $\theta_1,\dots \theta_k$ for the moment, statistically
decouple. Then
\begin{equation}
  \mathcal{L}_{\rm joint}\approx \mathcal{L}_{\rm joint}^*\exp
\left[ -\sum_{i=1}^k \frac{(\theta_i-\hat\theta_i)^2}{2\sigma_i^2}\right]\;,
\end{equation}
where $\sigma_i$ is the standard error of the estimate~$\hat\theta_i$.
Our integration of~$\mathcal{L}_{\rm joint}$ over the $k$-dimensional
parameter space yields $\mathcal{L}^*_{\rm joint}\prod_{i=1}^k \left(\sqrt{2\pi}
\sigma_i\right)$. When the sample is very large, each~$\sigma_i$ shrinks
like~$n^{-1/2}$, giving rise to an $n^{-k/2}$ factor, as seen in Eq.~(19),
which may equivalently be written
\begin{equation}
\exp(-{\rm BIC}/2)\equiv n^{-k/2} \mathcal{L}^*_{\rm joint}\;.
\end{equation}
This is the reason why many data sets being analyzed today in cosmology
utilize the BIC, which comes from Bayesian statistical inference, approximating
the outcome of a Bayesian procedure for deciding between models, with an
accuracy that improves with increasing~$n$. The precision typically becomes
very high when $n\gg k$, as we have in this paper \cite{Schwarz1978,Kass1995}.

\begin{table*}
\centering
\tiny
\caption{Model Selection Based on the Joint Analysis of $H(z)$, HII-Galaxy and Quasar-Core Data}\label{tab:jonit}
\begin{tabular}{lcccccccr}
\\
\hline\hline
\\
Model\qquad& $H_0$ & $\Omega_{\rm m}$ & $w_{\Lambda}$ & $\alpha$ & $\kappa$ & $\ell_{\rm core}$ & BIC & Prob\\
& (km s$^{-1}$ Mpc$^{-1}$) & & & & & (pc) & & $(\%)$ \\
\\
\hline
\\
$R_{\rm h}=ct$ & $62.336\pm 1.464$ & -- & -- & $4.889\pm 0.077$ & $33.491\pm 0.123$ & $11.709\pm 0.667$ & $611.1228$ & $96.7$ \\
$\Lambda$CDM & $67.013\pm 2.578$ & $0.347\pm 0.049$ & $-1$ (fixed) & $4.956\pm 0.085$ & $33.340\pm 0.143$ & $11.684\pm 0.659$ & $617.9843$ & $3.1$ \\
$w$CDM & $64.718\pm 3.088$ & $0.247\pm 0.108$ & $-0.693\pm 0.276$ & $4.922\pm 0.088$ & $33.415\pm 0.154$ & $11.700\pm 0.665$ & $623.4201$ & $0.2$ \\
\\
\hline\hline
\end{tabular}

\vskip 0.3in
\end{table*}

\section{Discussion}
\label{discussion}
A joint analysis allows us to obtain the best-fit results for more free parameters than
are available when using a single data set. For example, when using just the measurements
of $H(z)$, the optimization in $R_{\rm h}=ct$ yields just the value of $H_0$ \citep{Melia2013c}.
With the combined analysis, the value of $H_0$ must be rendered consistently with all three
data sets and the other variables, such as $\alpha$, $\kappa$, and $\ell_{\rm core}$. While the
additional variables offer added flexibility, the fact that critical parameters, such as $H_0$,
must optimize the fits of diverse data sets yields a tighter constraint on the model itself.
An important outcome of our work is a demonstration that $H_0$ evaluated in this fashion
is fully consistent (to within 1$\sigma$) with previously optimized values of the Hubble
constant based on the analysis of single data sets \citep{Melia2013c,Wei2016}. An additional
benefit of using a joint analysis, e.g., with $H(z)$ and HII galaxies, is that one can
separate the optimization of $\kappa$ and $H_0$, which is otherwise not possible using just
the HII galaxies by themselves \citep{Wei2016}.

Our analysis in this paper has used three very \mbox{different} kinds of measurement: the expansion
rate, the angular-diameter distance, and the luminosity distance, of diverse sources, so the
fact that the optimized results are self-consistent provides a more compelling conclusion
regarding the suitability and viability of the cosmological models. It is one thing
to demonstrate consistency with one observational signature at a time, but when three different
signatures are shown to be consistent with a single set of optimized parameters, principally
$H_0$, the result is much more compelling. The optimized value of $H_0$ in $R_{\rm h}=ct$ is
fully consistent with all 4 previous measurements of this constant
\citep{Melia2013c,Melia2015b,Wei2017,Melia2018}, i.e., $\sim 63$ km s$^{-1}$ Mpc$^{-1}$.

Likewise, the inferred values of $\Omega_{\rm m}$ and $w_{\rm de}$ are consistent with
their {\it Planck} optimizations to within 1$\sigma$ \citep{Planck2016}. Perhaps more
importantly, our inferred value of $H_0$, especially for $\Lambda$CDM, is much more
consistent with the {\it Planck} optimization than the locally measured value using
Cepheid variables \citep{Riess2011}. This reinforces the view held by some workers that
the difference between the high- and low-redshift measurements of $H_0$ is real and may
be due to the effects of a local Hubble bubble that alters the dynamics of expansion
within a distance of $\sim 300$ pc of us (corresponding to $z\lesssim 0.07$) \citep{Marra2013}.

\section{Conclusion}
\label{conclusion}
The results of this joint analysis affirm the viability of the $R_{\rm h}=ct$ cosmology.
The joint analysis favors this model with a significantly higher probability than the
standard model, or its variant $w$CDM. Should $R_{\rm h}=ct$ eventually emerge as the
correct cosmology, many important consequences would follow. Clearly, it represents
a significant step forward in helping us to understand the nature of dark energy. For
one thing, it could not be a cosmological constant. It would have to be dynamic, probably
representing new physics beyond the standard model of particle physics.

Most importantly, $R_{\rm h}=ct$ does not have a horizon problem, so it does not
need inflation. Given how difficult it has been to find a self-consistent and
complete model of inflation even after 30 years since its introduction, the
viability of $R_{\rm h}=ct$ may allow us to completely avoid this poorly understood
phenomenon altogether, because it simply may have never happened.

This paper has affirmed the benefits of carrying out model selection based on the
joint analysis of various, diverse kinds of data. This clearly calls for broadening 
the scope of this kind of work, with the inclusion of other observations at both
low- and high-redshifts. Ideally, one would wish to include all of the
available data in such a joint analysis, but this goal will require a great deal of
effort. As we have discussed in this manuscript, the work reported here makes the 
first serious attempt beyond simpler comparative studies based solely on single data 
sets. We have included three distinct kinds of measurement here. The difficulty with 
this procedure, of course, is finding data that are completely model independent or, 
at least, using them in a way that does not bias one model over another. We are 
currently working towards that goal, and constructive steps are being taken going 
forward. The outcome of this expanded analysis will be reported elsewhere. For the 
time being, however, the results presented in this manuscript constitute a significant
advance over previous model comparisons, promising even more compelling results
from future work.

\section*{Acknowledgments}
We are grateful to the anonymous referee for helping us improve the
content and presentation of this manuscript. We are also grateful for very helpful
discussions with Robert Maier concerning the use of the Bayes Information Criterion. 
This work was partially supported by the National Key R\&D Program of China 
(2017YFA0402600),the National Natural Science Foundation of China (Grants No. 11573006, 
11528306), the Fundamental Research Funds for the Central Universities and the Special 
Program for Applied Research on Super Computation of the NSFC-Guangdong Joint Fund 
(the second phase).


\begin{thebibliography}{90}
\bibitem[1]{Jimenez2002} R. {Jimenez} \& A. {Loeb}, Astrophys. J. {\bf 573} (2002) 37.
\bibitem[2]{Moresco2012} M. {Moresco}, L. {Verde} \& L. {Pozzetti et al.}, J. Cosmol. Astropart. Phys. {\bf 53} (2012) 012.
\bibitem[3]{Blake2012} C. {Blake}, S. {Brough} \& M. {Colless et al.}, Mon. Not. R. Astron. Soc. {\bf 425} (2012) 405.
\bibitem[4]{Melnick1987} J. {Melnick}, M. {Moles}, R. {Terlevich} \& J. M. {Garcia-Pelayo},
Mon. Not. R. Astron. Soc. {\bf 226} (1987) 849.
\bibitem[5]{Melnick1988} J. {Melnick}, R. {Terlevich} \& M. {Moles}, Mon. Not. R. Astron. Soc. {\bf 235} (1988) 297.
\bibitem[6]{Fuentes-Masip2000} O. {Fuentes-Masip}, C. {Mu{\~n}oz-Tu{\~n}{\'o}n}, H. O. {Casta{\~n}eda} \& G. {Tenorio-Tagle},
Astron. J. {\bf 120} (2000) 752.
\bibitem[7]{Melnick2000} J. {Melnick}, R. {Terlevich} \& E. {Terlevich}, Mon. Not. R. Astron. Soc. {\bf 311} (2000) 629.
\bibitem[8]{Bosch2002} G. {Bosch}, E. {Terlevich} \& R. {Terlevich}, Mon. Not. R. Astron. Soc. {\bf 329} (2002) 481.
\bibitem[9]{Telles2003} E. {Telles}, in Astronomical Society of the Pacific Conference Series,
  Vol. 297, Star Formation Through Time, ed. E.~{Perez}, R.~M. {Gonzalez
  Delgado}, \& G.~{Tenorio-Tagle} (2003) 143.
\bibitem[10]{Siegel2005} E. R. {Siegel}, R. {Guzm{\'a}n} \& J. P. {Gallego et al.},
Mon. Not. R. Astron. Soc. {\bf 356} (2005) 1117.
\bibitem[11]{Bordalo2011} V. {Bordalo} \& E. {Telles}, Astrophys. J. {\bf 735} (2011) 52.
\bibitem[12]{Plionis2011} M. {Plionis}, R. {Terlevich} \& S. {Basilakos et al.}, Mon. Not. R. Astron. Soc. {\bf 416} (2011) 2981.
\bibitem[13]{Mania2012} D. {Mania} \& B. {Ratra}, Phys. Lett. B {\bf 715} (2012) 9.
\bibitem[14]{Chavez2012} R. {Ch{\'a}vez}, E. {Terlevich} \& R. {Terlevich et al.}, Mon. Not. R. Astron. Soc. {\bf 425} (2012) L56.
\bibitem[15]{Terlevich2015} R. {Terlevich}, E. {Terlevich} \& J. {Melnick et al.}, Mon. Not. R. Astron. Soc. {\bf 451} (2015) 3001.
\bibitem[16]{Gurvits1999} L. I. {Gurvits}, K. I. {Kellermann} \& S. {Frey}, Astron. Astrophys. {\bf 342} (1999) 378.
\bibitem[17]{Cao2015} S. {Cao}, M. {Biesiada}, X. {Zheng} \& Z.-H. {Zhu} Astrophys. J. {\bf 806} (2015) 66.
\bibitem[18]{Cao2017a} S. {Cao}, M. {Biesiada} \& J. {Jackson et al.} J. Cosmol. Astropart. Phys. {\bf 2} (2017) 012.
\bibitem[19]{Cao2017b} S. {Cao}, X. {Zheng} \& M. {Biesiada et al.}, Astron. Astrophys. {\bf 606} (2017) A15.
\bibitem[20]{Wei2017} J.-J. {Wei}, F. {Melia} \& X.-F. {Wu}, Astrophys. J. {\bf 835} (2017) 270.
\bibitem[21]{Percival2009} W. J. {Percival} \& M. {White}, Mon. Not. R. Astron. Soc. {\bf 393} (2009) 297.
\bibitem[22]{Macaulay2013} E. {Macaulay}, I. K. {Wehus} \& H. K. {Eriksen}, Phys. Rev. Lett. {\bf 111} (2013) 161301.
\bibitem[23]{Pavlov2014} A. {Pavlov}, O. {Farooq} \& B. {Ratra}, Phys. Rev. D {\bf 90} (2014) 023006.
\bibitem[24]{Alam2016} S. {Alam}, S. {Ho} \& A. {Silvestri}, Mon. Not. R. Astron. Soc. {\bf 456} (2016) 3743.
\bibitem[25]{Peacock2001} J. A. {Peacock}, S. {Cole} \& P. {Norberg et al.}, Nature {\bf 410} (2001) 169.
\bibitem[26]{Guzzo2008} L. {Guzzo}, M. {Pierleoni} \& B. {Meneux et al.}, Nature {\bf 451} (2008) 541.
\bibitem[27]{Ross2007} N. P. {Ross}, J. {da {\^A}ngela} \& T. {Shanks et al.}, Mon. Not. R. Astron. Soc. {\bf 381} (2007) 573.
\bibitem[28]{daAngela2008} J. {da {\^A}ngela}, T. {Shanks} \& S.-M. {Croom et al.}, Mon. Not. R. Astron. Soc. {\bf 383} (2008) 565.
\bibitem[29]{Viel2004} M. {Viel}, M. G. {Haehnelt} \& V. {Springel}, Mon. Not. R. Astron. Soc. {\bf 354} (2004) 684.
\bibitem[30]{Davis2011} M. {Davis}, A. {Nusser} \& K. L. {Masters et al.}, Mon. Not. R. Astron. Soc. {\bf 413} (201) 2906.
\bibitem[31]{Hudson2012} M. J. {Hudson} \& S. J. {Turnbull}, Astrophys. J. Lett. {\bf 751} (2012) L30.
\bibitem[32]{Melia2014b} F. Melia, Astron. J. {\bf 147} (2014) 120.
\bibitem[33]{Melia2013a} F. Melia, Astrophys. J. {\bf 764} (2013) 72.
\bibitem[34]{Melia2015b} F. {Melia} \& T. M. {McClintock}, Astron. J. {\bf 150} (2015) 119.
\bibitem[35]{Steinhardt2016} C. L. {Steinhardt}, P. {Capak}, D. {Masters} \& J. S. {Speagle}, Astrophys. J. {\bf 824} (2016) 21.
\bibitem[36]{Yennapureddy2018} M. K. {Yennapureddy} \& F. {Melia}, Eur. Phys. J. C {\bf 78} (2018) 258.
\bibitem[37]{Melia2007} F. {Melia}, Mon. Not. R. Astron. Soc. {\bf 382} (2007) 1917.
\bibitem[38]{Melia2009} F. {Melia} \& M. {Abdelqader}, Int. J. Mod. Phys. D {\bf 18} (2009) 1889.
\bibitem[39]{Melia2012a} F. {Melia} \& A.S.H. {Shevchuk}, Mon. Not. R. Astron. Soc. {\bf 419} (2012) 2579.
\bibitem[40]{Melia2016b} F. Melia, Annals of Physics, in press (2019).
\bibitem[41]{Melia2017a} F. Melia, Mon. Not. R. Astron. Soc. {\bf 464} (2017) 1966.
\bibitem[42]{Melia2018} F. {Melia} \& M. K. {Yennapureddy}, J. Cosmol. Astropart. Phys. {\bf 2} (2018) 034.
\bibitem[43]{Melia2013c} F. {Melia} \& R. S. {Maier}, Mon. Not. R. Astron. Soc. {\bf 432} (2013) 2669.
\bibitem[44]{Cao2018} S.-L. Cao, X.-W. Duan, X.-L. Meng \& T.-J. Zhang, The European Physical Journal C {\bf 78} (2018) 313.
\bibitem[45]{Ruan2019} C.-Z. Ruan, F. Melia, Y. Chen \& T.-J. Zhang, Astrophys. J. {\bf 881} (2019) 137.
\bibitem[46]{Wei2016} J.-J. {Wei}, X.-F. {Wu} \& F. {Melia}, Mon. Not. R. Astron. Soc. {\bf 463} (2016) 1144.
\bibitem[47]{Erb2006} D. K. {Erb}, C. C. {Steidel} \& A. E. {Shapley et al.}, Astrophys. J. {\bf 646} (2006) 107.
\bibitem[48]{Hoyos2005} C. {Hoyos}, D. C. {Koo} \& A. C. {Phillips et al.}, Astrophys. J. Lett. {\bf 635} (2005) L21.
\bibitem[49]{Maseda2014} M. V. {Maseda}, A. {van der Wel} \& H.-W. {Rix et al.}, Astrophys. J. {\bf 791} (2014) 17.
\bibitem[50]{Masters2014} D. {Masters}, P. {McCarthy} \& B. {Siana et al.}, Astrophys. J. {\bf 785} (2014) 153.
\bibitem[51]{Chavez2014} R. {Ch{\'a}vez}, R. {Terlevich} \& E. {Terlevich et al.}, Mon. Not. R. Astron. Soc. {\bf 442} (2014) 3565.
\bibitem[52]{Chavez2016} R. {Ch{\'a}vez}, M. {Plionis}, S. {Basilakos}, R. {Terlevich}, E. {Terlevich}, J. {Melnick}, 
F. {Bresolin} \& A. L. {Gonz{\'a}lez-Mor{\'a}n}, Mon. Not. R. Astron. Soc. {\bf 462} (2016) 2431.
\bibitem[53]{Jackson2006} J. C. {Jackson} \& A. L. {Jannetta}, J. Cosmol. Astropart. Phys. {\bf 11} (2006) 002.
\bibitem[54]{Preston1985} R. A. {Preston}, D. D. {Morabito} \& J. G. {Williams et al.}, Astron. J. {\bf 90} (1985) 1599.
\bibitem[55]{Gurvits1994} L. I. {Gurvits}, Astrophys. J. {\bf 425} (1994) 442.
\bibitem[56]{Vishwakarma2001} R. G. {Vishwakarma}, Class. Quant. Grav. {\bf 18} (2001) 1159.
\bibitem[57]{Santos2008} R. C. {Santos} \& J.A.S. {Lima}, Phys. Rev. D {\bf 77} (2008) 083505.
\bibitem[58]{Melia2017b} F. Melia, Front. Phys. {\bf 12} (2017) 129802.
\bibitem[59]{Foreman-Mackey2013} D. {Foreman-Mackey}, D. W. {Hogg}, D. {Lang} \& J. {Goodman}, Publ. Astron. Soc. Pac.
{\bf 125} (2013) 306.
\bibitem[60]{Schwarz1978} G. Schwarz, Ann. Statist. {\bf 6} (1978) 461.
\bibitem[61]{Kass1995} R. E. Kass \& A. E. Raftery, J. Amer. Statist. Assoc. {\bf 90} (1995) 773.
\bibitem[62]{Planck2016} {Planck Collaboration et al.}, Astron. Astrophys. {\bf 594} (2016) A13.
\bibitem[63]{Riess2011} A. G. {Riess}, L. {Macri} \& S. {Casertano et al.}, Astrophys. J. {\bf 730} (2011) 119.
\bibitem[64]{Marra2013} V. Marra, L. Amendola, I. Sawicki \& W. Valkenburg, PRL {\bf 110} (2013) 241305.
\bibitem[65]{Zhang2014} C. {Zhang}, H. {Zhang} \& S. {Yuan et al.}, Res. Astron. Astrophys. {\bf 14} (2014) 1221.
\bibitem[66]{Jimenez2003} R. {Jimenez}, L. {Verde}, T. {Treu} \& D. {Stern}, Astrophys. J. {\bf 593} (2003) 622.
\bibitem[67]{Simon2005} J. {Simon}, L. {Verde} \& R. {Jimenez}, Phys. Rev. D {\bf 71} (2005) 123001.
\bibitem[68]{Moresco2016} M. {Moresco}, L. {Pozzetti} \& A. {Cimatti et al.}, J. Cosmol. Astropart. Phys. {\bf 5} (2016) 014.
\bibitem[69]{Stern2010} D. {Stern}, R. {Jimenez} \& L. {Verde, et al.}, J. Cosmol. Astropart. Phys. {\bf 2} (2010) 8.
\bibitem[70]{Moresco2015} M. {Moresco}, Mon. Not. R. Astron. Soc. Lett. {\bf 450} (2015) L16.
\end{thebibliography}
\end{document}